\documentclass[english,aps,prl,twocolumn,superscriptaddress,showpacs,reprint]{revtex4-1}
\usepackage{graphicx}
\usepackage{mathrsfs}
\usepackage[T1]{fontenc}
\usepackage{amsfonts}
\usepackage{amsmath}
\usepackage{amssymb}
\usepackage{subfigure}
\usepackage{bm}
\usepackage{verbatim}
\usepackage{color}
\usepackage{xcolor}
\usepackage{hyperref}
\usepackage{epsfig}
\usepackage{longtable}
\usepackage{footnote}
\usepackage{esint}
\newcommand{\upperRomannumeral}[1]{\uppercase\expandafter{\romannumeral#1}}

\begin{document}
\title{Mapping a fractional quantum Hall state to a fractional Chern insulator}
\author{Yinhan Zhang}
\affiliation{International Center for Quantum Materials, Peking University, Beijing 100871, China}
\affiliation{Collaborative Innovation Center of Quantum Matter, Beijing 100871, China}
\author{Junren Shi}\email{junrenshi@pku.edu.cn}
\affiliation{International Center for Quantum Materials, Peking University, Beijing 100871, China}
\affiliation{Collaborative Innovation Center of Quantum Matter, Beijing 100871, China}

\date{\today}

\begin{abstract}
We establish a variational principle for properly mapping a fractional quantum Hall (FQH) state to a fractional Chern insulator (FCI).  We find that the mapping has a gauge freedom which could generate a class of FCI ground state wave functions appropriate for different forms of interactions. Therefore, the gauge should be fixed by a variational principle that minimizes the interaction energy of the FCI model. For a soft and isotropic electron-electron interaction, the principle leads to a gauge coinciding with that for maximally localized \emph{two-dimensional} projected Wannier functions of a Landau level.
\end{abstract}
\maketitle

There has been a surge of interest in the realization and exploration of exotic topological states in condensed matter systems, such as quantum anomalous Hall insulators~\cite{Zhang2013,Haldane-model} and topological insulators~\cite{Konig2007,Kane2010a}. Recently, theoretical studies reveal a class of topological systems called fractional Chern insulator (FCI), which exhibits fractionally quantized Hall effect in a periodic system with a fractionally filled flat Chern band (FCB)~\cite{Sheng2011,Tang2011,Sun2011,Neupert2011}. While most of the studies so far are numerical simulations~\cite{Bergholtz2013}, it is still a current interest to find a systematic way to construct its ground-state wave function (GSWF) analytically~\cite{Parameswaran2011,Qi2011,Roy2014,Murthy2011,Murthy2012,Lu2012,McGreevy2012}. 

Qi proposes that the GSWF could be constructed from ordinary factional quantum Hall (FQH) wave functions (e.g. Laughlin's wave function)~\cite{Qi2011}. Specifically, the FQH wave function is expanded in a set of single-particle bases of the lowest Landau level (LLL), which are substituted with a set of bases constructed from the FCB. The approach is attractive because the ordinary FQH system, i.e., a two-dimensional (2D) electron gas subjected to a strong magnetic field with a fractionally filled LLL, is unique in that we know how to construct its wave function and determine its various properties~\cite{Laughlin1983, Jains, Haldane1983, Halperin1984}. However, the approach has arbitrariness in how the basis sets are mapped between the LLL and the FCB. Actually, Qi adopts a mapping between one-dimensional maximally localized Wannier functions~\cite{Qi2011}, while Wu {\it et al.} develop a more intricate mapping which is believed to provide a better GSWF~\cite{Wu2012b, Wu2013}. Although these approaches achieve successes to a certain extent, choosing a rule for the basis mapping is still largely empirical. It is desirable to establish a general principle to determine the best way to map a FQH wave function to its FCI counterpart.     

In this Letter, we establish a variational principle for properly mapping a GSWF of a FQH state in a continuous space to a FCI in a discretized lattice. We survey the most general form of the mapping by explicitly constructing two-dimensional localized Wannier functions, using which one could map a multi-band continuous model to a FCB tight-binding (TB) model. We observe that there exists a gauge freedom in constructing the Wannier functions, and the different gauges will result in TB models of the same kinetic part but very different forms of interactions. As a result, the gauge freedom should be treated as variational parameters of the GSWF, and be fixed by a variational principle minimizing the interaction energy of the FCI model. For a soft and isotropic electron-electron interaction, we show that the principle leads to a gauge coinciding with that for maximally localized \emph{two-dimensional} projected Wannier funtions of a LLL.

\begin{figure}[tb]
\centering
\includegraphics[width=0.5\columnwidth]{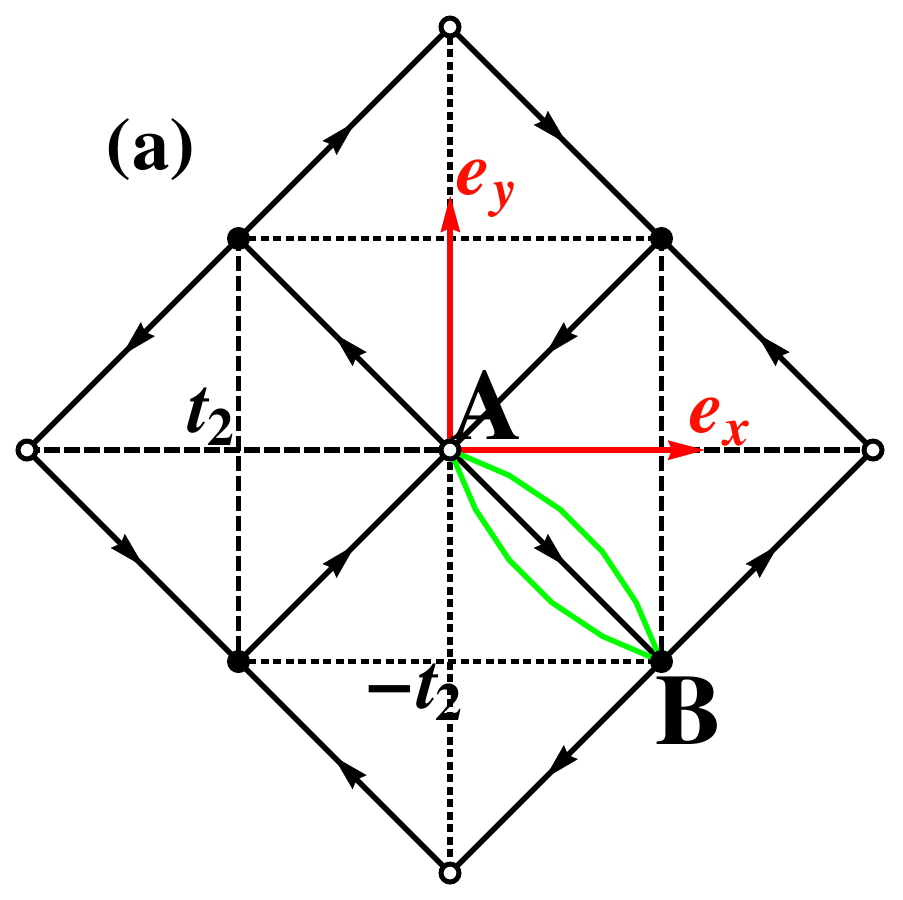}
\includegraphics[width=0.5\columnwidth]{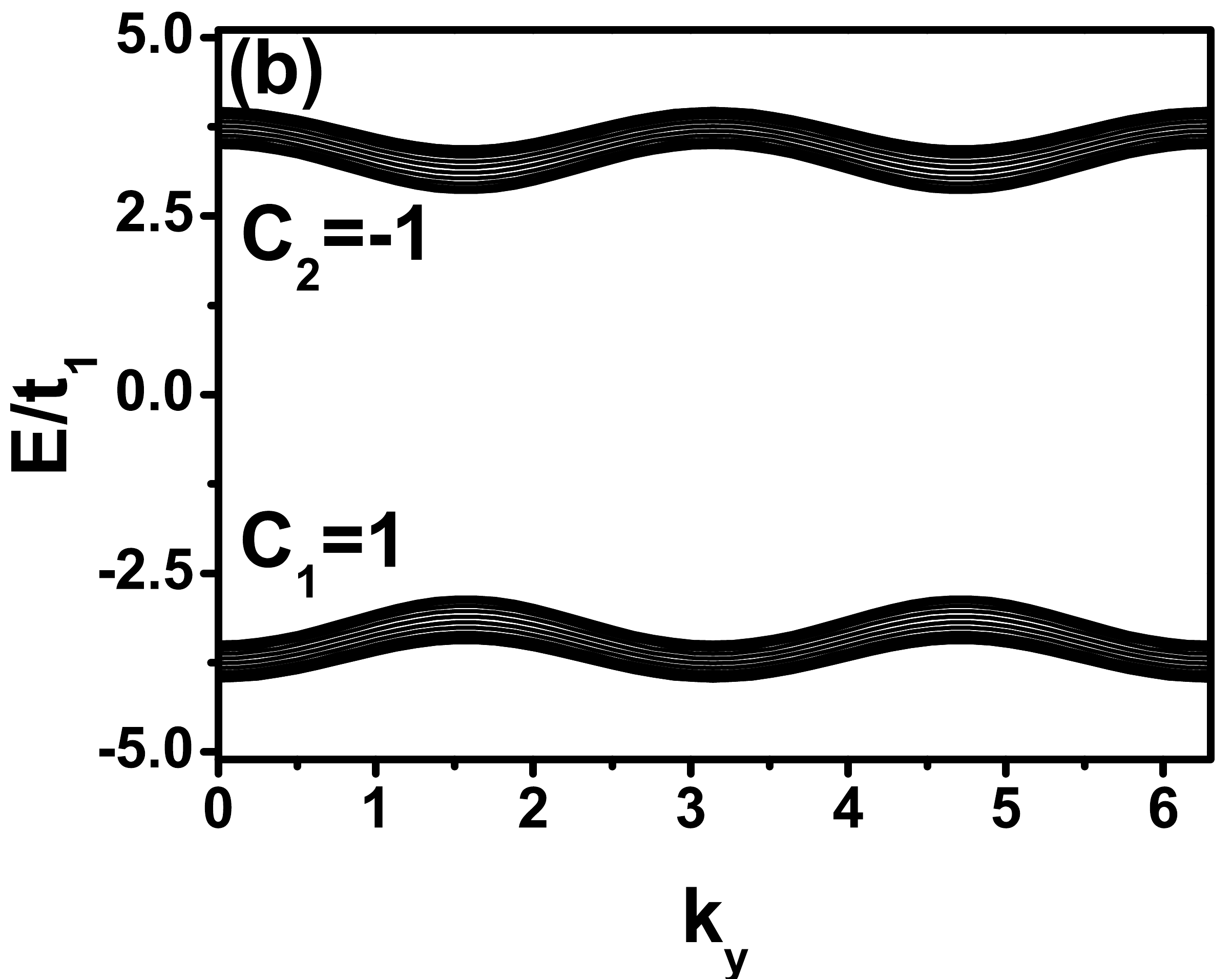}
\caption{(color online) The chiral $\pi$-flux model~\cite{Neupert2011}: (a) The schematic diagram describes a square lattice with a chiral $\pi$-flux. Each unit cell contains two sites (A and B).  The nearest-next (NN) hopping amplitudes are $t_{1}\exp(i\pi/4)$ in the direction of the arrow (solid lines) and the next-nearest-next (NNN) hopping amplitudes are $t_{2}$ and $-t_{2}$ along the dashed and dotted lines, respectively; (b) The band structure when $t_{1}/t_{2}=\sqrt{2}$. }
\label{fig:TBModel}
\end{figure}   

We consider a general 2D tight-binding model, 
\begin{equation}
\hat{H}=\sum_{\bm k}\hat{a}^{\dagger}_{\bm k}h(\bm k)\hat{a}_{\bm k }+\hat{h}_{int},
\label{lattice-model}
\end{equation}
where $h(\bm k)$ is a $\mathcal{N}\times\mathcal{N}$ hermitian matrix, $\hat{a}(\bm k)$ ($\hat{a}^{\dagger}(\bm k)$) is an $\mathcal{N}$-component annihilation (creation) operator at the quasi-momentum $\bm k$ defined in a Brillouin zone. We can diagonalize $h(\bm k)$ to give rise to a set of bands with dispersions $\epsilon_n(\bm k)$ and eigen-vectors $u_{n}(\bm k)$, $n=1\dots\mathcal{N}$. We are particularly interested in a class of FCB systems, which have one or more isolated bands with nearly vanishing dispersions and nonzero Chern numbers. When a FCB is fractionally filled, the interaction $\hat{h}_{int}$ may drive the system to a FQH phase, as that happens in a fractionally filled Landau level, resulting in a FCI.  The interaction is usually assumed to be $\hat{h}_{int} = \sum_{\bm i,\bm j,\tau_1,\tau_2} V^{\tau_1\tau_2}(\bm R_i - \bm R_j) \hat{n}_{\bm i \tau_1} \hat{n}_{\bm j \tau_2}$, where $\hat{n}_{\bm i \tau}$  is an electron number operator at the $\tau$-th site in the $\bm i$-th unit cell with the lattice vector $\bm R_i$. We further assume that the energy scale of the interaction is much larger than the bandwidth of the FCB, but much smaller than band gaps separating the FCB from the other bands.  In this case, it is sufficient to consider a projection of the interaction to the FCB: $\hat{h}^{\mathrm{P}}_{int}=(1/N)\sum_{\bm k_1, \bm k_2, \bm q}M(\bm k_1,\bm k_2; \bm q) \hat{\rho}_{\bm k_1,\bm q} \hat{\rho}_{\bm k_2, -\bm q}$, where $\hat{\rho}_{\bm k,\bm q}\equiv \hat{d}_{\bm k+\bm q}^\dagger \hat{d}_{\bm k}$, $\hat{d}^\dagger_{\bm k}$ and $\hat{d}_{\bm k}$ are creation and annihilation operators in the FCB, respectively, $N$ is the total number of unit cells, and the interaction matrix element is:
\begin{multline}
M(\bm k_1,\bm k_2; \bm q) = \sum_{\tau_1,\tau_2}V^{\tau_{1}\tau_{2}}_{\bm q} \\
u_{1,\tau_{1}}^{\ast}(\bm{k}_1 + \bm q) u_{1,\tau_{1}}(\bm{k}_1) 
u_{1,\tau_{2}}^{\ast}(\bm{k}_2 - \bm q) u_{1,\tau_{2}}(\bm{k}_2),
\label{Mint}
\end{multline}
where $u_{1,\tau}$ is $\tau$-th component of the eigen-vector for the fractionally filled FCB, which is assumed to have index $n=1$, and $V_{\bm q}^{\tau_1\tau_2} =  \sum_{\bm R} V^{\tau_1\tau_2}(\bm R) \exp (-i\bm q \cdot \bm R)$. 

To obtain a GSWF for the FCI model Eq.~(\ref{lattice-model}) from its FQH counterpart, it is crucial to establish a proper mapping between the LLL to the FCB.  Indeed, if we could ever map a parent Hamiltonian of a FQH wave function (e.g., Laughlin'wave function) to the FCI model, we would readily obtain an \emph{exact} FCI GSWF by applying the same mapping to the FQH wave function.  Unfortunately, this is not always possible by mapping the single-particle bases only. Nevertheless, we can optimize the approach by requiring that:  (\upperRomannumeral{1}) the mapping should map a continuous model containing a LLL into a TB model that has the exactly same form of the kinetic part of Eq.~(\ref{lattice-model}), and map the LLL to the FCB;  (\upperRomannumeral{2}) among all possible mappings satisfying (\upperRomannumeral{1}), we should pick the best one based on a general principle instead of {\it ad hoc} assumptions.    

By explicitly constructing a continuous model and mapping it to the TB model, we will show that the requirement (\upperRomannumeral{1}) can be fulfilled. The construction also makes clear that there exists a gauge freedom that cannot be fixed by the requirement (\upperRomannumeral{1}) only.  As a result, the most general form of the FCI-GSWF using the basis substituting approach can be written as, 
\begin{equation}
|\Psi[\theta(\bm k)]\rangle=\sum_{\{\bm k\}}e^{{i\sum_{\bm k}\theta(\bm k)}}|\{\bm k\} \rangle{}\left[_{L}\langle \{\bm k\} |\Phi\rangle_{L} \right],
\label{FCIGWF}
\end{equation}
where  $|\Phi\rangle_{L}$ is an ordinary FQH many-body wave function defined on a torus, ${}_{L}\langle\{\bm k\}|$ is a Slater determinant of a set of magnetic Bloch bases in the LLL with 2D quasi-momentums $\{\bm k\}$, and $|\{\bm k\}\rangle$ is a Slater determinant with the same set of quasi-momentums $\{\bm k\}$, albeit constructed from Bloch bases of the FCB model. In the formalism, the mapping is achieved by substituting each magnetic Bloch basis of the LLL with a Bloch basis of the FCB of the same quasi-momentum. Different choices of $\{\theta(\bm k)\}$ will map the FQH wave function to different FCI wave functions. It is easy to show that both Qi's and Wu {\it et al.}'s formalism could be cast to the form. 

Equation (\ref{FCIGWF}) can also be considered as a trial wave function for the FCI with variational parameters $\{\theta(\bm k)\}$. To fix $\{\theta(\bm k)\}$, we can apply the variational principle of the ground state energy. It is equivalent to minimize functional:
\begin{multline}
E_{int}[\theta(\bm k)]= \frac{1}{N}\sum_{\bm k_1, \bm k_2, \bm q} M(\bm k_1,\bm k_2; \bm q)
\Pi(\bm k_1 - \bm k_2;\bm q) \\
\times e^{-i\left[\theta(\bm k_1 + \bm q)-\theta(\bm k_1)+\theta(\bm k_2 -\bm q)-\theta(\bm k_2)\right]}, 
\label{interaction}
\end{multline}
where $\Pi(\bm k_1 -\bm k_2;\bm q) = \langle \hat{c}^\dagger_{\bm k_1+\bm q} \hat{c}^\dagger_{\bm k_2-\bm q}\hat{c}_{\bm k_2}\hat{c}_{\bm k_1}\rangle_{FQH}$, and $\hat{c}_{\bm k}$ ($\hat{c}_{\bm k}^\dagger$) is annihilation (creation) operator of a magnetic Bloch state of the LLL at quasi-momentum $\bm k$. The correlation function can be derived from the two-particle reduced density matrix of the FQH state (see Eq.~(\ref{correlation})).  With Eq.~(\ref{interaction}), we have a general variational principle to fix $\theta(\bm k)$, and fulfill the requirement (\upperRomannumeral{2}).  

The variational problem can be analytically solved for a limiting case when the electron-electron interaction is soft and isotropic, i.e., $V^{\tau_1\tau_2}_{\bm q} = V(q)$, and $V(q)\ne 0$ only for $qa\ll 1$, where $a$ is the lattice constant of the system.  In this case, $M(\bm k_1, \bm k_2;\bm q) \approx V(q) \exp[i(\bm A_{\bm k_1} - \bm A_{{\bm k}_2})\cdot \bm q]$, where $\bm A_{\bm k} = i \left\langle u_{1}(\bm k) \left| \partial_{\bm k} u_{1}(\bm k)\right. \right\rangle$ is the Berry connection of the FCB. We find that $\theta(\bm k)$ can be determined by a "Coulomb gauge" condition:
\begin{equation}
\bm \nabla_{\bm k}\cdot \left(-\bm\nabla_{\bm k} \theta(\bm k) + \bm A_{\bm k} - \bm A_{\bm k}^{L} \right) = 0,
\label{CoulombGauge}
\end{equation}
where $\bm A^L_{\bm k}$ is the Berry connection for the magnetic Bloch band of the LLL. It can be shown that the gauge is equivalent to that for constructing maximally localized 2D Wannier orbits projected to the FCB~\cite{Marzari1997,supplementary}.

To obtain these results, we first explicitly construct a continuous model and map it to the TB model Eq.~(\ref{lattice-model}). We define a Hilbert space spanned by a set of continuous-space bases $\{\psi_{n\bm k}(\bm r),\,n=1\dots\mathcal{N}\}$, and $\psi_{n\bm k}(\bm r)=\varphi_{n\bm k}(\bm r)\hat{\chi}_{n}$, where  $\varphi_{n\bm k}(\bm r)$ is a Bloch wave function defined in the real space with the same periodicity of the FCI model, and $\hat{\chi}_{n}$ denotes a $\mathcal{N}$-component pseudo-spin standard basis with $\langle\chi_n |\chi_{n^\prime}\rangle = \delta_{nn^\prime}$.  We require that $\varphi_{n\bm k}(\bm r)$ should give rise to the same Chern number $\mathcal{C}_n$ as its FCI counterpart $u_n(\bm k)$, and be an eigenstate of a hamiltonian $\hat{h}_n$ with the same dispersion $\epsilon_n(\bm k)$.  The hamiltonian of the total system can then be written as $\hat{\mathcal{T}}={\rm{diag}}(\hat{h}_{1},...,\hat{h}_{\mathcal{N}})$, which is the continuous model to be mapped to the TB model. It is easy to see that the choice of $\{\varphi_{n\bm k}(\bm r),\, \hat{h}_n \}$ is not unique, and one can freely choose a convenient one. For instance, for the FCB $n=1$, if $\mathcal{C}_1=\pm 1$, one can choose $\varphi_{1\bm k}(\bm r)$ to be the the magnetic Bloch wave function of the LLL, and $\hat{h}_1$ the hamiltonian for an electron moving in a 2D plane subjected to a uniform perpendicular magnetic field. The choice reproduces both the flat dispersion and the Chern number of the FCB~\cite{footnote}, and is the most convenient one for mapping a FQH wave function defined on the LLL.

We map the continuous model to the TB model Eq.~(\ref{lattice-model}) by constructing a set of 2D localized Wannier functions.  To do that, we need a set of Bloch wave functions that are continuous and periodic functions of the quasi-momentum in the Brillouin zone.  However, due to the topological obstruction of the nonzero Chern number, $\psi_{n\bm k}(\bm r)$ cannot be both continuous and periodic~\cite{Thouless1984}. Nevertheless, because we can freely choose a phase factor for $\psi_{n\bm k}(\bm r)$ at each $\bm k$, we can regularize wave functions $\{\psi_{n\bm k}(\bm r)\}$, following a procedure prescripted in Ref.~\onlinecite{Thouless1984} (see also Supplemental Material~\cite{supplementary}), to make them continuous inside the Brillouin zone, and satisfying quasi-periodicity conditions:
\begin{gather}
\psi_{n\bm{k}+\bm{K}_{1}}(\bm{r})=\psi_{n\bm{k}}(\bm{r}),\notag\\
\psi_{n\bm{k}+\bm{K}_{2}}(\bm{r})=\psi_{n\bm{k}}(\bm{r})\exp(i\bm k\cdot \bm a_{1}\mathcal{C}_{n}),
\label{BC1}
\end{gather}
where $\bm a_{1(2)}$ and $\bm K_{1(2)}$ are primitive lattice vectors in real and reciprocal spaces, respectively.  We can then define a new set of Bloch wave functions:
\begin{equation}
\phi_{\tau\bm k}(\bm r)=\sum_{n=1}^{\mathcal{N}}\psi_{n \bm k}(\bm r)u^\ast_{n,\tau}(\bm k),\,\,\,\tau=1\dots\mathcal{N},
\label{U2}
\end{equation}
where $u_n(\bm k)$ should also be regularized using the same procedure as that for $\psi_{n\bm k}(\bm r)$.  As a result, it is also continuous inside the Brillouin zone and satisfies the same quasi-periodicity conditions as Eq.~(\ref{BC1}). It is easy to verify that $\phi_{\tau\bm k}(\bm r)$ is both continuous and periodic in the Brillouin zone.  Using $\phi_{\tau\bm k}(\bm r)$, we can construct a set of 2D localized Wannier functions $w_{\tau}(\bm r;\bm R)=(1/\sqrt{N})\sum_{\bm k}\phi_{\tau\bm k}(\bm r)\exp(-i\bm k\cdot\bm R)$. Moreover, it is easy to show that
\begin{align}
\langle \phi_{\tau_1\bm k}|\hat{\mathcal{T}}|\phi_{\tau_2\bm k}\rangle=
\sum_n u_{n,\tau_1}(\bm k)\epsilon_n(\bm k)u_{n,\tau_2}^\ast(\bm k) = h_{\tau_1\tau_2}(\bm k),
\label{HH}
\end{align}
i.e., using the $\{\phi_{\tau\bm k}(\bm r)\}$ as bases, the continuous model $\hat{\mathcal{T}}$ is mapped to the kinetic part $h(\bm k)$ of the FCI model Eq.~(\ref{lattice-model}). With this, we explicitly demonstrate that we can construct and map a continuous model containing a LLL into a TB model that has the exactly same form of the kinetic part of Eq.~(\ref{lattice-model}).

It now becomes clear that there exists a gauge freedom $\theta(\bm k)$ indicated in Eq.~(\ref{FCIGWF}), because we can replace $u_{n}(\bm k)\rightarrow u_{n}(\bm k)e^{i\theta_{n}(\bm k)}$ (or equivalently, $\psi_{n\bm k}\rightarrow \psi_{n\bm k}(\bm k)e^{-i\theta_{n}(\bm k)}$ ), where $\theta_{n}(\bm k)$ ($\theta(\bm k) \equiv \theta_1(\bm k)$) is a continuous and periodic (modulo $2\pi$) function of $\bm k$, to obtain another valid mapping. Different gauge choices do not affect the mapping to the kinetic part of the FCI model, as that is apparent from Eq.~(\ref{HH}). However, they will lead to different spatial distributions of the Wannier functions. As a result, the mapping of the interaction part will sensitively depend on the gauge choice.

To demonstrate the dependence, we explicitly discretize an interaction $\hat{V}(\bm r, \bm r^\prime)=-V\nabla^{2}_{\bm r}\delta(\bm r-\bm r^\prime)$ ($V>0$), which is the pseudo-potential for  Laughlin's state at filling factor $\nu=1/3$~\cite{Trugman1994}.  For the TB model, we consider a chiral $\pi$-flux model~\cite{Neupert2011}, as detailed in Fig.~\ref{fig:TBModel}. It is sufficient to assume that the interaction resides only inside the LLL, since only the interaction projected into the FCB is relevant~\cite{Bergholtz2013,Roy2011}. In general, the mapping will generate an interaction more complicated than that assumed in $\hat{h}_{int}$.  However, for our demonstration purpose, we are only concerned with the interaction matrix elements appeared in $\hat{h}_{int}$: 
\begin{multline}
\tilde{V}^{\tau_{1}\tau_{2}}(\bm{R}_{1}-\bm{R}_{2})= V \int d\bm{r} \\
 \left(\bm\nabla \left| w^{(1)}_{\tau_{1}}(\bm{r};\bm{R}_{1}) \right|^2 \right) 
\cdot 
\left(\bm\nabla \left| w^{(1)}_{\tau_{2}}(\bm{r};\bm{R}_{2}) \right|^2 \right) 
\end{multline}  
where $w^{(1)}_{\tau}(\bm r;\bm R) =(1/\sqrt{N})\sum_{\bm k}\varphi_{1\bm k}(\bm r)u^\ast_{1,\tau}(\bm k)\exp(-i\theta(\bm k)-i\bm k\cdot\bm R )$ is Wannier function projected to the FCB.   
     
\begin{figure}[tbh!]
\centering
\includegraphics[width=0.49\columnwidth]{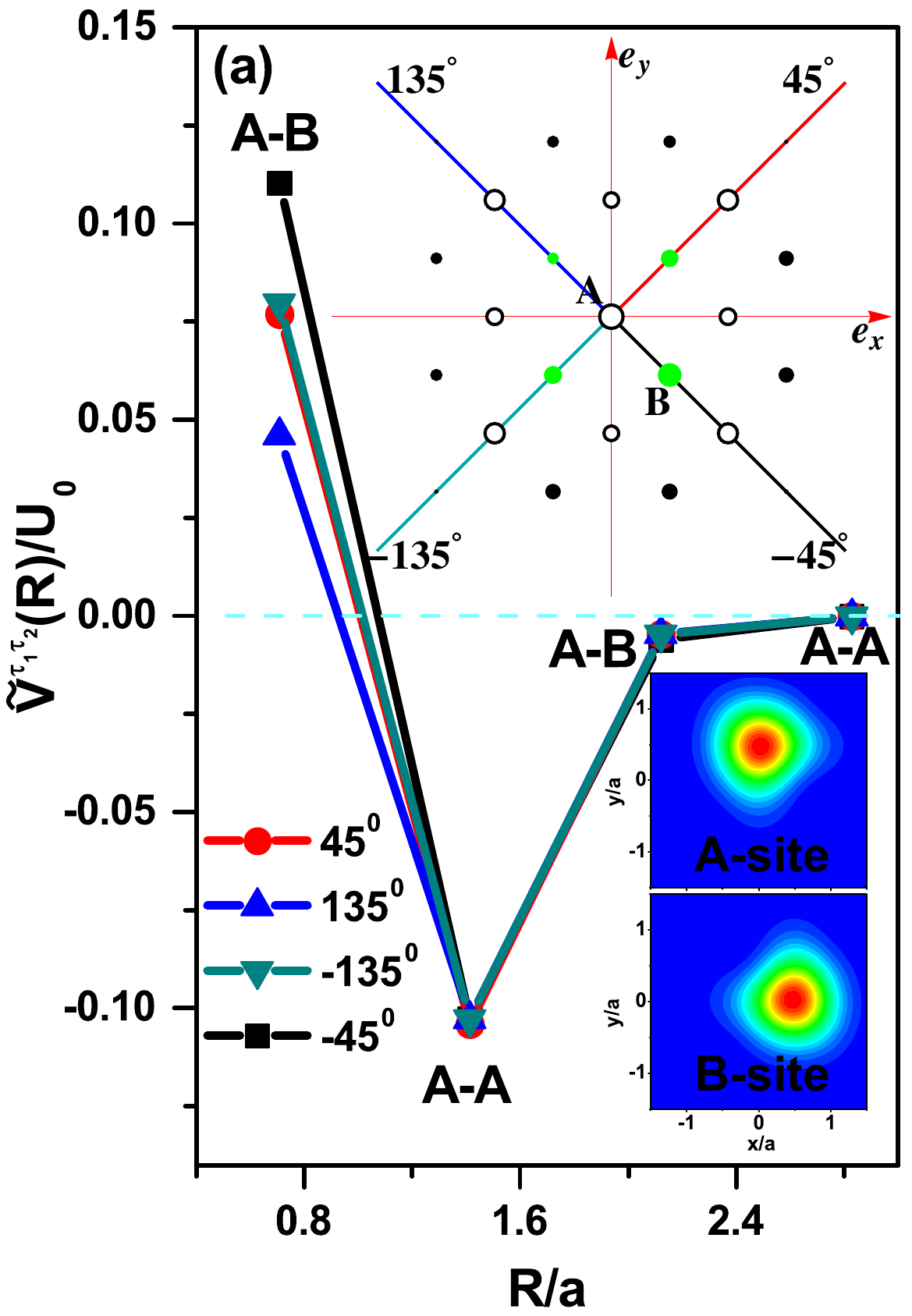}
\includegraphics[width=0.49\columnwidth]{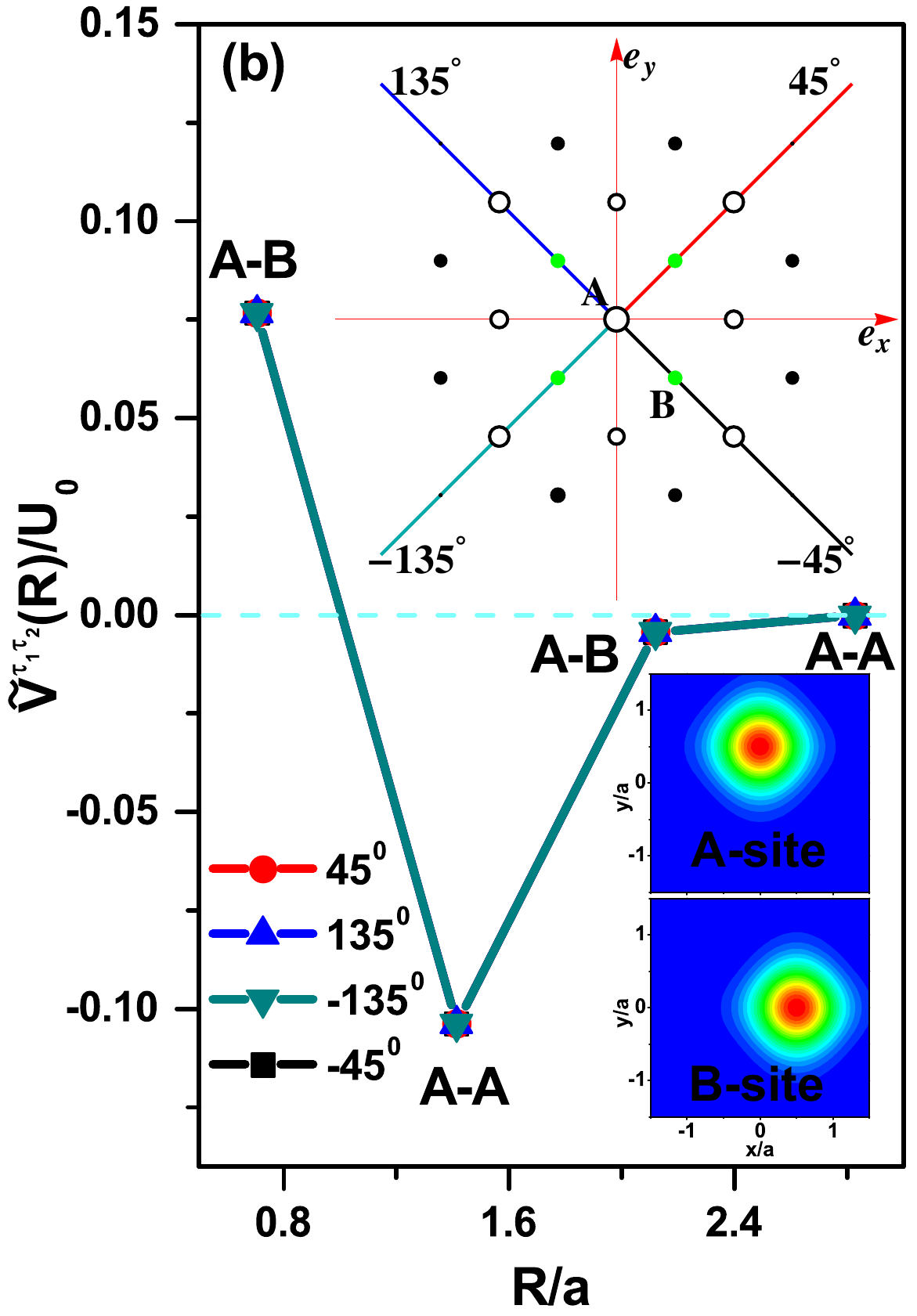}
\caption{(color online) $\tilde{V}^{\tau_1\tau_2}(\bm R)$ mapped from the pseudo-potential for Laughlin's state at $\nu=1/3$,  as a function of distance $R/a$ along four different directions $\pm45^{\circ}$ and $\pm 135^{\circ}$ for (a) $\theta(\bm k)=0 $, and (b) symmetric gauge~\cite{supplementary}. $U_{0}=0.7524V$ is the strength of the on-site interaction. The inset contour plots in (a) or (b) show spatial distributions of the two Wannier functions for site-A and site-B, respectively.}
\label{fig:WannierHubbardU}
\end{figure}

We first consider the case of $\theta(\bm k)=0$. Such a gauge actually corresponds to the one used in Qi's proposal~\cite{supplementary,Qi2011,Barkeshi2012}. We immediately observe that neither spatial distributions of the Wannier functions nor the mapped interaction $\tilde{V}^{\tau_1\tau_2}(\bm R)$ have four-fold symmetry of the original FCI model, as shown in Fig.~\ref{fig:WannierHubbardU}(a). It suggests that if such a mapping could ever generate an exact FCI-GSWF, the interaction of the FCI model should be highly anisotropic. 

The lack of the symmetry is not intrinsic. Actually, it is easy to find a gauge $\theta(\bm k)$ to make the density distribution of $w^{(1)}_{\tau}(\bm r;\bm R)$ four-fould symmetric~\cite{supplementary}. This is shown in Fig.~\ref{fig:WannierHubbardU}(b). For a typical FCI model which usually has an interaction with the full symmetry of the underlying lattice, one would expect that the latter choice of gauge  could generate a better GSWF for the FCI model.

The above observation suggests that different gauges would give rise to FCI-GSWFs appropriate for different forms of interaction.  As a result, the gauge should be fixed by the interaction part for a given FCI model. To do that, we note that the mapping of the Bloch bases corresponds to a substitution of the creation (annihilation) operator of the FCB by that of the LLL:
\begin{align}
\hat{d}_{\bm k}\rightarrow \hat{c}_{\bm k},\, \,\,
\hat{d}_{\bm k}^\dagger\rightarrow \hat{c}_{\bm k}^\dagger.
\end{align}   
To take account of the gauge freedom, we apply the substitution $u_{1}(\bm k)\rightarrow u_{1}(\bm k)e^{i\theta(\bm k)}$ in Eq.~(\ref{Mint}), resulting in an extra phase factor for the interacting matrix element $M(\bm k_1,\bm k_2,\bm q)$.  By applying the variational principle for the ground state energy, and noting that the gauge freedom has no impact on the kinetic energy, it is straightforward to obtain the variational functional Eq.~(\ref{interaction}).  

To evaluate the functional $E_{int}[\theta(\bm k)]$, one needs to determine the FQH correlation function $\Pi(\bm k_1 -\bm k_2,\bm q)$, which is related to the two-particle reduced density matrix $\rho_{2}(\bm r_{1}^\prime,\bm r_{2}^\prime;\bm r_{1},\bm r_{2})$ for the FQH state on a torus by
\begin{multline}
\Pi(\bm k_1 -\bm k_2,\bm q) =
\int\prod_{i=1}^{2}d\bm{r}_{i}d\bm{r}_{i}^\prime  
\rho_{2}(\bm{r}^\prime_{1},\bm{r}^\prime_{2};\bm{r}_{1},\bm{r}_{2})  \\ 
 \times \varphi_{1\bm{k}_{1}+\bm q}(\bm{r}_{1}^\prime)\varphi_{1\bm{k}_{2}-\bm q}(\bm{r}_{2}^\prime) 
\varphi_{1\bm{k}_{1}}^{\ast}(\bm{r}_{1})\varphi_{1\bm{k}_{2}}^{\ast}(\bm{r}_{2}).
\label{correlation}
\end{multline}
For a FQH state of the LLL, $\rho_{2}(\bm r_{1}^\prime,\bm r_{2}^\prime;\bm r_{1},\bm r_{2})$ can be fully determined by its pair correlation functions~\cite{Vignale2005}. Using a general expression of the pair correlation function developed in Ref.~\onlinecite{MacDonald1988}, we can obtain a closed expression for $\Pi(\bm k_1 -\bm k_2,\bm q)$~\cite{supplementary}. We find that, in general, $\Pi(\bm k_1 -\bm k_2,\bm q)$ can be decomposed into:
\begin{multline}
\Pi(\bm k_1 -\bm k_2,\bm q) = \nu^2\left(\delta_{\bm q,0} - \delta_{\bm k_1-\bm k_2+\bm q,0}\right)\\
+ e^{-i(\bm A^L_{\bm k_1 +\bm q}-\bm A^L_{\bm k_2})\cdot\bm q}\Pi^\prime(\bm k_1 -\bm k_2,\bm q),
\end{multline}  
where $\nu$ is the filling factor of the FQH state, $\bm A^L_{\bm k}= -\mathcal{C}_1 \bm a_1 (\bm k\cdot \bm a_2 )/2\pi$ is the Berry connection of the LLL, and $\Pi^\prime(\bm k_1 -\bm k_2,\bm q)$ is a periodic function of $\bm k_1 - \bm k_2$.

For the soft and isotropic electron-electron interaction, we can assume that $\bm q$ is small when evaluating Eq.~(\ref{interaction}). To the first order of $\langle q^2 \rangle \equiv \int d\bm q V(q)q^2$, we find that:
\begin{multline}
E_{int} \propto \langle q^2\rangle \sum_{\bm k_1,\bm k_2}\Pi_0(\bm k_1-\bm k_2) 
\left(-\bm\nabla_{\bm k_1}\theta(\bm k_1)  + \Delta\bm A_{\bm k_1}\right)\\
\times\left(-\bm\nabla_{\bm k_2} \theta(\bm k_2) + \Delta\bm A_{\bm k_1}\right)
+\mathcal{O}\left(\langle q^4\rangle\right) ,
\end{multline} 
where $\Pi_0(\bm k_1 -\bm k_2) \equiv \Pi^\prime(\bm k_1 -\bm k_2,\bm 0) - \delta_{\bm k_1, \bm k_2}\sum_{\bm k} \Pi^\prime(\bm k,\bm 0)$, $\Delta\bm A_{\bm k} \equiv \bm A_{\bm k} - \bm A^L_{\bm k}$, and terms independent of $\theta(\bm k)$ are ignored.  The functional can be rewritten as $E_{int} \propto \langle q^2\rangle \sum_{\bm R} \tilde{\Pi}_0(\bm R) |i\bm R \tilde{\theta}(\bm R) + \Delta \tilde{\bm A}_{\bm R} |^2 $, where $\tilde{\theta}(\bm R)$, $\tilde{\Pi}_0(\bm R)$, $\Delta \tilde{\bm A}_{\bm R}$ are Fourier transformations of $\theta(\bm k)$, $\Pi_0(\bm k)$, $\Delta \bm A_{\bm k}$ with $\tilde{f}(\bm R) = \sum_{\bm k} f(\bm k) \exp(i\bm k\cdot \bm R)$, respectively. Noting that $\tilde{\Pi}_0(\bm R) \geq 0$ for all $\bm R$, which is tested to be true for $\nu =1/3$ and $\nu=1/5$ Laughlin's states~\cite{supplementary}, we conclude that the minimization of $E_{int}$ is equivalent to minimizing $|i\bm R \tilde{\theta}(\bm R) + \Delta \tilde{\bm A}_{\bm R}|^2$ for all $\bm R$, which immediately leads to the "Coulomb gauge" condition Eq.~({\ref{CoulombGauge}}). Moreover, we can show that the gauge is exactly the one which minimizes the total extent of the projected Wannier functions $ r^2_w \equiv  \sum_{\tau}\int d\bm r |w^{(1)}_\tau(\bm r, \bm 0) |^2 \bm r^2$. It is interesting to observe that in this limiting case the proper choice of the gauge is fully determined by the topology of the FCB, independent of the FQH state to be mapped. However, for the more general form of the interaction, we expect that the dependence would emerge.

To conclude, we have established a variational principle to fix the gauge freedom when constructing ground-state wave functions for fractional Chern insulators by mappings from fractional quantum Hall states. Our theory is built upon the general variational principle of the ground state energy, free of {\it ad hoc} assumptions. It suggests that the proper mapping is not solely determined by quantities of the band topology such as Berry connection and Berry curvature, in contrast to the assumption of most of previous theoretical efforts. In general, it also depends on the form of the interaction and the FQH state to be mapped. All these factors should be considered in a unified variational principle minimizing Eq.~(\ref{interaction}). Moreover, with Eq.~(\ref{FCIGWF}), we have a understanding on the general structure of FCI-GSWF. This could be useful if we would want to develop an scheme (e.g., a counterpart of the composite fermion picture) to directly construct FCI-GSWF at the level of the TB model. Finally, we show that the gauge for the maximally localized (projected) Wannier orbits, which are widely adopted in first-principles calculations~\cite{Marzari1997}, could naturally emerge as the gauge choice for soft and isotropic electron-electron interactions in this particular problem.

We gratefully acknowledge the useful discussions with Kai Sun, Kun Yang, Zixiang Hu and Hua Chen, and the supports from the National Basic Research Program of China (973 Program) Grant Nos. 2012CB921304 and 2015CB921101, and from National Science Foundation of China (NSFC) Grant No. 11325416.


\newpage

\vskip 15 cm
\begin{widetext}
\setcounter{figure}{0}
\setcounter{equation}{0}
\setcounter{section}{0}

\renewcommand\thefigure{S\arabic{figure}}
\renewcommand\theequation{S\arabic{equation}}

\section{Supplementary information for "Mapping a fractional quantum Hall state to a fractional Chern insulator''}

\vskip 0.5 cm
In this supplemental material, we show the procedure to obtain the initial gauge used in the main text, and its relation to the gauge in Qi's proposal. We also show how a symmetric gauge used in Fig.~2(b) for the chiral $\pi$-flux model could be constructed. We will obtain a closed expression for $\Pi(\bm k-\bm k^\prime,\bm q)$. Finally, we show that "Coulomb gauge" condition is equivalent to that for constructing maximally localized projected Wannier functions.    

\subsection{The procedure for regularizing U(1) phases of Bloch wave functions}
Thouless shows that the phase factors of the Bloch wave functions can be assigned by imposing the constraints~\cite{Thouless1984}:
\begin{align}
\langle\psi_{0,k_{2}}|\partial_{k_{2}}|\psi_{0,k_{2}}\rangle=0, \\
\langle\psi_{k_{1},k_{2}}|\partial_{k_{1}}|\psi_{k_{1},k_{2}}\rangle=0,
\end{align} 
where $k_{1(2)}\in [0,1]$ denotes two components of the quasi-momentum, and $\bm k=k_{1}\bm K_{1}+k_{2}\bm K_{2}$.  The resulting Bloch wave functions will be continuous inside the Brillouin zone, and satisfying the quasi-periodicity conditions:
\begin{align}
\bm \psi_{\bm k+\bm K_{2}}(\bm r)=\exp(i\delta_{0})\bm \psi_{\bm k}(\bm r),\\
\bm \psi_{\bm k+\bm K_{1}}(\bm r)=\exp(i\delta(k_{2}))\bm \psi_{\bm k}(\bm r),
\end{align}
and $\delta(k_{2}+1)=\delta(k_{2})-2\pi \mathcal{C}$, where $\mathcal{C}$ is the Chern number of the band.

We can make a further transformation: 
\begin{align}
\psi_{\bm k}(\bm r) \rightarrow \psi_{\bm k}(\bm r)\exp\left(-\frac{i}{2\pi}\bm k \cdot (\bm a_{2}\delta_{0}+\bm a_{1}\delta(k_{2}))\right).
\end{align} 
It is easy to verify that the wave functions satisfies the quasi-periodic condition Eq.~(6) of the main text.  This is the initial gauge used in the main text

For a rectangular Brillouin zone, the gauge obtained from the above procedure can be explicitly written as, 
\begin{eqnarray}
\psi(\bm k)&\rightarrow \exp\left(-i\delta(k_{y})\frac{k_{x}}{K_{x}}-i\frac{\delta_{0}k_{y}}{K_{y}}\right)\exp(i\int_{0}^{k_{x}}dp_{x}a_{x}(p_{x},k_{y}))\exp(i\int_{0}^{k_{y}} d p_{y} a_{y}(0,p_{y}))\psi(\bm k),
\label{TG}
\end{eqnarray}
where $\bm a(\bm k)$ is the Berry connection of the initial Bloch wave functions, and $\delta_{0}=\int_{0}^{K_{y}}dp_{y}a_{y}(0,p_{y})$, $\delta(k_y)=\int_{0}^{K_{x}}dp_{x}a_{x}(p_{x},k_y)$. It is exactly Qi's generic gauge (See Ref. [30] in the main text, the gauge used in Ref. [10] can be obtained by setting $a_{y}=0$). 

The magnetic Bloch wave function of the LLL in the initial gauge for a rectangular Brillouin zone can be written as, 
\begin{equation}
\varphi_{\bm k}(\bm r)=\frac{1}{\sqrt{N_{x}\pi^{1/2}l_{M}L_{y}}}\sum_{l\in \mathbb{Z}}\exp\left(-iak_{x}l+i(k_{y}+lK_{y})y\right)\exp\left(-\frac{(x+(k_{y}+K_{y}l)l^2_M)^2}{2l^2_{M}}\right),
\label{MBloch}
\end{equation}
where $l_{M}$ is the magnetic length, and $L_{y}$ is the length of the system along $y$-direction. For given periodicity, $2\pi l_M^2 = s_u$, where $s_u$ is the area of a unit cell.

\subsection{Symmetric Gauge}

In this section, we show how to find a symmetric gauge for the chiral $\pi$-flux model. Because of $c_{2z}$ symmetry of the TB model, the Bloch wave functions of $\bm k$ and $-\bm k$ is related by,
\begin{align}
u(-\bm k)=\exp(i\vartheta(\bm k)) \left [\begin{array}{cc}
\exp(-i\bm{k}\cdot \bm a_{2}) & 0\\
0 & \exp(-i\bm{k}\cdot\bm{a}_{1})
\end{array}\right] u(\bm k)
\end{align}  
where $\vartheta(\bm k)$ can be determined numerically and is found to be periodic function of the momentum. To obtain symmetric Wanneir functions, we eliminate $\vartheta(\bm k)$ by a gauge transformation $u(\bm k)\rightarrow u(\bm k)\exp(i\theta(\bm k))$, and
\begin{equation}
\theta(\bm k)=-\frac{\vartheta(\bm k)}{2}.
\label{theta}
\end{equation}
By noting that $\varphi_{\bm k}(-\bm r) = \varphi_{-\bm k}(\bm r)$,  it is easy to verify that the Wannier functions have the following symmetry:  
\begin{equation}
P_{c^{-1}_{2z}}[w_{1}(\bm r;\bm R),w_{2}(\bm r;\bm R)]= [w_{1}(\bm r;c_{2z} \bm R-\bm a_{2}),w_{2}(\bm r;c_{2z} \bm R-\bm a_{1})]. 
\label{SymmetryWannier} 
\end{equation}
The symmetry is sufficient to make the interacting matrix elements symmetric.

\subsection{The calculation of the closed expression for $\Pi(\bm k-\bm k^\prime,\bm q)$}    

In this section, we obtain a closed form for the correlation function $\Pi(\bm k-\bm k^\prime,\bm q)$.  The two-particle density matrix is defined as 
\begin{equation}
\rho_{2}(\bm r_{1},\bm r_{2};\bm r^\prime_{1},\bm r^{\prime}_{2}) = N_{e}(N_{e}-1)\int \prod_{j=3}^{N_{e}} d \bm r_{j}\Phi^{\ast}_{\nu}(\bm r_{1},\bm r_{2},\bm r_{3}...\bm r_{N_{e}})\Phi_{\nu}(\bm r^\prime_{1},\bm r^{\prime}_{2}, \bm r_{3},...,\bm r_{N_{e}}).
\end{equation}
The analytic property of the many-body wave function in the LLL allows us relating the two-particle density matrix to the pair correlation function (see Ref.~[32] of the main text). This is because a many body wave function in the LLL can always be written as (for the gauge $A_x=-By$, $A_y = 0$),
\begin{equation}
\Phi_{\nu}(\bm r_{1},...,\bm r_{N_{e}})=\exp\left(-\sum_{j=1}^{N_{e}}\frac{x^{2}_{j}}{2l^2_{M}}\right)G(z_{1},...,z_{N_{e}}),
\end{equation}     
where the $G(z_{1},...,z_{N})$ is an holomorphic function about $z_{j}=y_{j}+ix_{j}$ ($j=1,...,N_{e}$). As a result, the two-particle reduced density matrix must have the form, 
\begin{equation}
\rho_{2}(\bm r_{1},\bm r_{2}; \bm r^\prime_{1},\bm r^{\prime}_{2})=\exp\left(-\frac{x^{2}_{1}+x^2_2+x^{\prime 2}_{1}+x^{\prime 2}_{2}}{2l^{2}_M}\right)F(\bar z_{1},\bar z_{2};z^\prime_{1},z^\prime_{2}),
\label{rho2}
\end{equation}   
where $F(\bar z_{1}, \bar z_{2};z^\prime_{1},z^\prime_{2})$ is analytic for each argument. Because the diagonal element of reduced density matrix  is proportional to the pair correlation function $g(r)$, we have
\begin{equation}
F(\bar z_{1},\bar z_{2};z_{1},z_{2})=n^{2}g(r)\exp\left(-\frac{(z_{1}-\bar z_{1})^2+(z_{2}-\bar z_{2})^{2}}{4l^{2}_M}\right),
\end{equation}   
with $r^2 \equiv (z_{1}-z_{2})(\bar z_{1}-\bar z_{2})$. The full form of $F(\bar z_{1}, \bar z_{2};z^\prime_{1},z^\prime_{2})$ can be obtained by a substitution $z_{1 (2)}\rightarrow z^\prime_{1 (2)}$ in the above expression. After substituting it back into Eq.~(\ref{rho2}), we obtain the two-particle reduced density matrix:  
\begin{align}  
\rho_{2}(\bm r_{1},\bm r_{2}; \bm r^\prime_{1},\bm r^{\prime}_{2})=n^2\exp\left (-\sum_{j=1,2}\frac{(x_{j}-x_{j}^{\prime})^{2}+(y_{j}^{\prime}-y_{j})^{2}+2i(x_{j}^{\prime}+x_{j})(y_{j}^{\prime}-y_{j})}{4l_{B}^{2}}\right) g(r),
\label{rhodef} 
\end{align}
with $r^2 \equiv (z_{1}^\prime-z_{2}^\prime)(\bar z_{1}-\bar z_{2})$.

The pair distribution function $g(\bm r)$ can in general be written as the form~\cite{Girvin1986},
\begin{equation}
g(r)=1-\exp(-\frac{r^{2}}{2l^{2}_M})+2\sum_{k=0}^{\infty}\frac{c_{2k+1}}{(2k+1)!}(\frac{r^{2}}{4l^{2}_M})^{2k+1}\exp(-\frac{r^2}{4l^{2}_M}) 
\end{equation}
with the coefficients $c_{2k+1}$ tabulated in Ref.~\cite{Girvin1986} for filling factors $\nu=1/3$ and $\nu=1/5$.

To calculate the correlation function $\Pi(\bm k-\bm k^\prime, \bm q)$, we substitute Eqs.~(\ref{MBloch}) and (\ref{rhodef}) into Eq.~(11) of the main text.  We introduce an auxiliary pair distribution function 
\begin{equation}
g(\bm r,\alpha)=\exp\left(-\frac{(1+\alpha)r^{2}}{4l^2_M}\right),
\end{equation}
and calculate the corresponding auxiliary correlation function.  After collecting all terms, we find that,  
\begin{equation}
\Pi^\prime(\bm k, \bm q)=2\sum^{\infty}_{k=0}\frac{c_{2k+1}}{(2k+1)!}(-\frac{\partial}{\partial \alpha})^{2k+1}\Pi^\prime(\bm k,\bm q;\alpha)|_{\alpha=0}
\label{PiPi}
\end{equation}
and,
\begin{align}
\Pi^\prime&(\bm k, \bm q;\alpha)=\nu^{2}\frac{4\pi l^2_M}{L_{x}L_{y}}\frac{1}{1+\alpha}\exp\left(-\pi\frac{1-\alpha}{1+\alpha}\frac{q^2}{K_{x}K_y}\right)
 \notag\\
 &\vartheta_{3}\left(\frac{k_{x}+q_{x}}{K_{x}}-i\frac{1-\alpha}{1+\alpha}\frac{q_{y}}{K_{x}}\right | \left.i\frac{1-\alpha}{1+\alpha}\frac{K_{y}}{K_{x}}\right)\vartheta_{3}\left(\frac{k_{y}+q_{y}}{K_{y}} +i\frac{1-\alpha}{1+\alpha}\frac{q_{x}}{K_{y}}\right |\left. i\frac{1-\alpha}{1+\alpha}\frac{K_{x}}{K_{y}}\right), 
\label{closedform}
\end{align} 
where the Jacobi theta function is defined as $\vartheta_{3}(z|\tau)=\sum_{n\in \mathbb{Z}} \exp(i\pi\tau n^2)\exp(2\pi i n z)$. 

Alternatively, the correlation function can be written as,
\begin{align}
\Pi^\prime(\bm k, \bm q) &= \frac{1}{N}\sum_{\bm R} \tilde{\Pi}^\prime(\bm R, \bm q) e^{-i\bm k\cdot \bm R} \\
\tilde \Pi^{\prime}(\bm R,\bm q) &= 2\nu^2\exp\left(-i\bm q\cdot \bm R
-\frac{1}{2l_M^2}\left(\bm{R} - \hat{z}\times\bm{q} l_M^2\right)^{2}\right)\sum^\infty_{k=0}
c_{2k+1}L_{2k+1}\left(\frac{1}{l_M^2}\left(\bm{R} - \hat{z}\times\bm{q} l_M^2\right)^{2} \right),
\end{align}
where $L_n(x)$ is a Laguerre function.

It is easy to see that,
\begin{align}
\tilde{\Pi}_{0}(\bm R)\equiv\tilde{\Pi}^\prime(\bm R,0)-\tilde{\Pi}^\prime(0,0),
\end{align}
which is shown in Fig.~\ref{fig:Fourier} for the filling factors $\nu=1/3$ and $\nu=1/5$. Obviously, $\tilde\Pi_{0}(\bm R)$ is always positive for these filling factors.

\begin{figure}[tbh!]
\centering
\includegraphics[width=0.5\columnwidth]{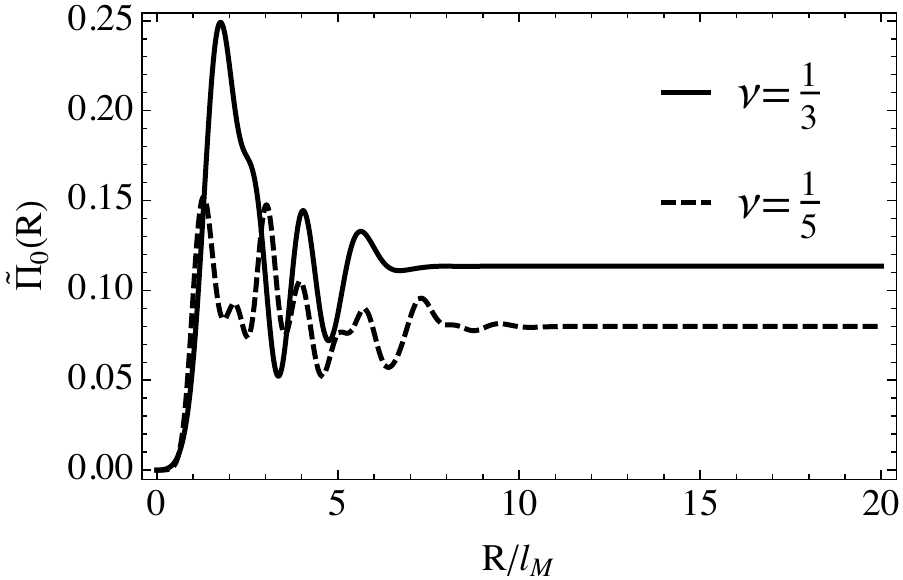}
\caption{ $\tilde{\Pi}_{0}(\bm R)$ for FQH liquid as a function of $R/l_M$ at $\nu=1/3$ (solid line) and $\nu=1/5$ (dashed line). }
\label{fig:Fourier}
\end{figure} 

\subsection{Gauge condition for the maximally localized projected Wannier functions}

In this section, we will show that the "Coulumb gauge" condition will minimize the extent of the projected Wannier functions. It is easy to show that:
\begin{align}
r^2_{w}=\sum_{\tau}\left[\frac{1}{N}\sum_{\bm k}\langle\partial_{\bm k}\tilde{\phi}_{\tau \bm k}|\cdot |\partial_{\bm k}\tilde{\phi}_{\tau \bm k}\rangle\right]
\end{align}
where $\tilde{\phi}_{\tau \bm k}(\bm r)=\exp(-i\bm{k}\cdot\bm r)\varphi_{1\bm k}(\bm r)u^\ast_{1,\tau}(\bm k)$ is periodic part of Bloch functions $\phi_{\tau \bm k}(\bm r)$ projected into FCB. Under a gauge transformation $u_{1\tau}(\bm k)\rightarrow u_{1\tau}(\bm k)\exp(i\theta(\bm k))$, the extent of Wannier function becomes 
\begin{equation}
r^2_{w}\rightarrow r^2_{w}+\sum_{\bm k}\left((\bm A^{L}_{\bm k}-\bm A_{\bm k}+\partial_{\bm k}\theta(\bm k))^2-(\bm A^{L}_{\bm k}-\bm A_{\bm k})^2\right).
\end{equation}
Minimizing $r^2_w$ over $\theta(\bm k)$ leads to condition: 
\begin{equation}
\nabla_{\bm k}\cdot (\bm A_{\bm k}-\bm A^{L}_{\bm k}-\nabla_{\bm k}\theta(\bm k))=0,       
\end{equation}
which is exactly the "Coulumb gauge" condition Eq.~(5) of the main text.

\end{widetext}


\begin{thebibliography}{99}

\bibitem{Zhang2013} C. -Z. Chang {\it et al.}, Science \textbf{340}, 6129 (2013)

\bibitem{Haldane-model} F.D.M. Haldane, Phys. Rev. Lett \textbf{61}, 2015 (1988).

\bibitem{Kane2010a} M. Z. Hasan and C. L. Kane, Rev. Mod. Phys. \textbf{82}, 3045 (2010). 

\bibitem{Konig2007} M. König, {\it et al.}, Science \textbf{318}, 766 (2013) 2007

\bibitem{Sheng2011} D. N. Sheng, Zheng-Cheng Gu, Kai Sun and L. Sheng, Nature commun. \textbf{2}, 389 (2011). 

\bibitem{Tang2011} E. Tang, J.-W. Mei, X.-G. Wen, Phys. Rev. Lett. \textbf{106}, 236802 (2011).

\bibitem{Sun2011} K. Sun, Z.-C. Gu, H. Katsura, S. Das Sarma, Phys. Rev. Lett. \textbf{106}, 236803 (2011).

\bibitem{Neupert2011} T. Neupert, L. Santos, C. Chamon, C. Mudry, Phys. Rev. Lett. \textbf{106}, 236804 (2011). 

\bibitem{Bergholtz2013} Emil J. Bergholtz and Zhao Liu, Int. J. Mod. Phys. B 27, 1330017 (2013).

\bibitem{Qi2011} X.-L. Qi, Phys. Rev. Lett. \textbf{107}, 126803 (2011).

\bibitem{Parameswaran2011} S. A. Parameswaran, Rahul Roy, and Shivaji L. Sondhi, Phys. Rev. B 85, 241308(R) (2011).

\bibitem{Roy2014} Rahul Roy, Phys. Rev. B \textbf{90}, 165139 (2014).

\bibitem{Murthy2011} G. Murthy and R. Shankar, arXiv:1108.5501.

\bibitem{Murthy2012} G. Murthy and R. Shankar, Phys. Rev. B \textbf{86}, 195146 (2012).

\bibitem{Lu2012} Y. M. Lu and Y. Ran, Phys. Rev. B \textbf{85}, 165134 (2012)

\bibitem{McGreevy2012} J. McGreevy, B. Swingle and K. -A. Tran, Phys. Rev. B \textbf{85} 125105 (2012)

\bibitem{Laughlin1983} R. B. Laughlin, Phys. Rev. Lett. \textbf{50}, 1395 (1983)

\bibitem{Jains} J. K. Jain, 2007, {\it Composite Fermions} (Cambridge University Press, 2007)

\bibitem{Haldane1983} F. D. M. Haldane, Phys. Rev. Lett. \textbf{51}, 605 (1983).

\bibitem{Halperin1984} B. I. Halperin, Phys. Rev. Lett. \textbf{52}, 1583 (1984).

\bibitem{Wu2012b} Y.-L. Wu, B. A. Bernevig, and N. Regnault, Phys. Rev. B \textbf{86}, 085129 (2012).

\bibitem{Wu2013} Y.-L. Wu, N. Regnault, and B. Andrei. Bernevig, Phys. Rev. Lett. \textbf{110}, 106802 (2013).    

\bibitem{Marzari1997} Nicola Marzari and David Vanderbilt, Phys. Rev. B \textbf{56} 12847 (1997)

\bibitem{supplementary} See Supplemental Material at [url], which includes Refs.~\cite{Thouless1984,Girvin1986}

\bibitem{Thouless1984} D. J. Thouless, J. Phys. C: Solid State Phys. \textbf{17}, L325 (1984).

\bibitem{Girvin1986} Girvin, S. M., MacDonald, A. H., and Platzman, P. M., Phys, Rev. B \textbf{33}, 2481 (1986). 

\bibitem{footnote} The small non-flatness of $\epsilon_1(\bm k)$ can be considered as a perturbation to the FCI phase, and is ignored here.


\bibitem{Trugman1994} S. A. Trugman and S. Kivelson, Phys. Rev. B \textbf{50}, 17199 (1994).

\bibitem{Roy2011} R. Roy and S. L. Sondhi, Physics 4, 46 (2011).

\bibitem{Barkeshi2012} M. Barkeshli and X.-L. Qi, Phys. Rev. X \textbf{2}, 031013 (2012). 

\bibitem{Vignale2005} Gabriele F. Giuliani and Giovanni Vignale  {\it Quantum Theory of the Electron Liquid} (Cambridge University Press, 2005)

\bibitem{MacDonald1988} MacDonald, A. H. and Girvin, S. M., Phys. Rev. B \textbf{38}, 6295 (1988).























\end{thebibliography}

\begin{thebibliography}{References}

\bibitem{Thouless1984} D. J. Thouless, J. Phys. C: Solid State Phys. \textbf{17}, L325 (1984).

\bibitem{Girvin1986} Girvin, S. M., MacDonald, A. H., and Platzman, P. M., Phys, Rev. B \textbf{33}, 2481 (1986). 


\end{thebibliography}
\end{document}